\documentclass[preprint,12pt]{elsarticle}

\usepackage{amssymb}
\usepackage{algorithm}
\usepackage{algpseudocode}
\usepackage{subfigure}
\usepackage{booktabs}
\usepackage{multirow}
\usepackage{amsmath}
\usepackage[misc]{ifsym}
\usepackage{fancyhdr}

\journal{Computer Speech \& Language}

\begin{document}

\begin{frontmatter}



\title{Adversarial example devastation and detection on speech recognition system by adding random noise}


\author{Mingyu Dong}
\author{Diqun Yan\corref{cor1}}\ead{yandiqun@nbu.edu.cn}
\author{Rangding Wang}
\address{College of Information Science and Engineering, Ningbo University, Ningbo Zhejiang, China}
\cortext[cor1]{Corresponding author}


\begin{abstract}

An automatic speech recognition (ASR) system based on a deep neural network is vulnerable to  attack by an adversarial example, especially if the command-dependent ASR fails. A defense method against adversarial examples is proposed to improve the robustness and security of the ASR system. We propose an algorithm of devastation and detection on adversarial examples that can attack   current advanced ASR systems. We choose an  advanced text-  and command-dependent ASR system as our target, generating adversarial examples by an optimization-based attack on text-dependent ASR and the GA-based algorithm on command-dependent ASR. The  method is based on input transformation of   adversarial examples. Different random intensities and kinds of noise are added to   adversarial examples to devastate the perturbation previously added to   normal examples. Experimental results show that the method performs well. For the devastation of examples, the original speech similarity  after adding noise can reach 99.68\%, the similarity of  adversarial examples can reach zero, and the detection rate of   adversarial examples can reach 94\%.
\end{abstract}




\end{frontmatter}


\section{Introduction}
\label{intro}
Deep neural network technology has been   used in many fields \cite{Ref1,Ref2}, and related  security problems   have become increasingly prominent, among which the adversarial example \cite{Ref3} is   of    great concern. In 2014, Szegedy et al. \cite{Ref4} found that the deep neural network (DNN) showed high vulnerability to image examples with specific perturbation, including  adversarial perturbation. Their study   is of great significance to explain the principle of deep learning, and has promoted the development of security attack and defense based on deep learning.

Automatic speech recognition (ASR)   \cite{Ref add1} has been used  for intelligent speech assistance and vehicle speech control systems, helping users control and connect to services through simple speech. These systems are vulnerable to attack by adversarial examples. Vaidya et al. \cite{Ref5} first proposed a method to generate speech adversarial examples. The ASR system recognized errors when it adjusted the parameters extracted by Mel-frequency cepstral coefficients (MFCCs) \cite{Ref6}. Carlini et al. \cite{Ref7} extended the work   to hide malicious commands in speech,   gave a more detailed description and analysis of the scene of speech adversarial examples, and gave   methods of white-    and black-box attacks. Alzantot et al. \cite{Ref8}   applied genetic algorithms to black-box attacks and successfully attacked command-dependent ASR systems. Yuan et al. \cite{Ref9} hid speech commands in music. Cisse et al. \cite{Ref10} proposed a more flexible attack method, which can be applied to different models. To attack the end-to-end ASR model, the method required the loss of the target command and the current prediction result, and found an adversarial example through optimization. Iter et al. \cite{Ref11} generated adversarial examples based on the trained WaveNet \cite{Ref12} model. This method is mainly based on the fast gradient sign algorithm (FGSM) \cite{Ref13}.

Some work \cite{Ref7,Ref8} is aimed at command-dependent ASR systems, causing speech to be misclassified. However, the application is   text-based in   most practical scenes. Because the generated text is indefinite in length, there will be some problems   calculating the loss between the output   and   target texts. Carlini \cite{Ref14} introduced CTC loss \cite{Ref15} to the adversarial example of speech recognition to solve this problem. Speech adversarial example research has also extended from command-  to text-dependent speech,   including the following work. Carlini \cite{Ref14} solved the problem of gradient backpropagation in the MFCC computing process, because many ASR systems do not directly accept   original speech, and need to extract its MFCC coefficients. The method can   use any   original speech to  generate adversarial examples. The   success rate for white-box attacks is close to 100\%. However, this method has several problems: (1) it does not consider   playback in the real scene, and the efficiency of generating speech is low; (2) it takes nearly an hour to generate adversarial examples; (3) the perturbation is large; (4) the method cannot be applied to a black-box; and (5) the adversarial examples generated by the optimization of a model are only effective for that model.

The following work considered the above shortcomings \cite{Ref14}. Large perturbation is mainly related to the measurement of the difference between  generated  and  original examples. Qin et al. \cite{Ref16} adopted a psychoacoustic model to redesign the loss function, so that the adversarial example was closer to the original   in hearing. Carlini's algorithm converged with difficulty, resulting in low attack efficiency. Schonherr et al. \cite{Ref17} improved the algorithm so that adversarial examples could be generated in several minutes. Liu et al. \cite{Ref18} also improved the efficiency of adversarial example generation. Taori et al. \cite{Ref19} combined a genetic algorithm with gradient estimation   to solve the adversarial example problem of a text-based speech recognition system in a black-box scene, but only obtained 35\%  accuracy.

The defense of adversarial examples  helps researchers find and fix   security loopholes that may occur in  ASR systems based on deep learning. Defense methods   include devastation and detection of adversarial examples. Devastation    causes an adversarial example to lose its attack ability without affecting the context of normal examples. The   detection strategy   determines whether an example is   adversarial, and those are discarded.

For the devastation strategy, Latif et al. \cite{Ref20} used a generative adversarial network (GAN) to denoise   input examples, causing adversarial examples to lose their attack ability. Yang et al. \cite{Ref21} used U-Net to enhance the input data to invalidate   adversarial examples. Sun et al. \cite{Ref22} took the adversarial examples generated by various  algorithms as extended datasets to retrain the network. Experimental results showed that the network model after such training could better resist   adversarial examples. Samizade et al. \cite{Ref23} designed a convolutional neural network (CNN)-based method to detect adversarial examples. Rajaratnam et al. \cite{Ref24} detected adversarial examples by adding random noise to different frequency bands of speech. Rajaratnam et al. \cite{Ref25}   proposed  to detect adversarial speech examples by comparing the differences between adversarial   and normal examples in feature space.

There is   a simple and effective method to simultaneously devastate and detect, which only needs to modify the input speech examples. In addition to using time-dependence to detect adversarial examples, Yang et al. \cite{Ref26} found some effective modification methods to defend against adversarial examples from   defense methods in the image field, which include local smoothing, downsampling, and re-quantization. Kwon et al. \cite{Ref27}   used a number of speech modification methods,  including low-pass filtering, 8-bit re-quantization, and mute processing, to defend the adversarial examples. Methods based on speech modification   cause   loss of   information of normal speech examples, which is usually unacceptable.

This work  researches  adversarial example defense to improve  the security and robustness of ASR. The attack success rate of adversarial examples should be reduced as much as possible while ensuring the recognition accuracy of normal examples. To this end, we propose a speech adversarial example defense algorithm based on the addition of random noise. Through a large number of experiments, we find that after adding a specific random noise to an adversarial example, its perturbation  will be transformed to the summary of the original perturbation and random noise. Due to the influence of random noise, the original perturbation will be devastated and lose its particularity, and the adversarial example will   lose its attack ability. Experimental results show that this method can effectively defend against adversarial examples of the two kinds of ASR systems, with a better defense effect   than other methods.

The rest of this paper is organized as follows. Section 2  introduces   work related to   classic adversarial example generation. Section 3 describes our proposed devastation and detection method for adversarial examples. The setting of the experiment,  devastation of   adversarial examples, and  experimental results are discussed in section 4.  Section 5   summarizes our work.

\section{Related work}
\label{sec:2}
ASR systems can be categorized as either text- or command-dependent, which respectively recognize   input speech as a text sequence or command tag. ASR systems employ different attack methods. We  introduce the two typical speech adversarial example attack methods. optimization-based method (OPT) is based on gradient optimization \cite{Ref14} on a text-based ASR system, and another   method is based on a genetic algorithm (GA) \cite{Ref8} for a command-dependent system.

\subsection{OPT method}
\label{sec:2-1}
Given a speech example $x$, a perturbation $\delta$ can be constructed that is almost imperceptible to   human hearing, but     $x+\delta$  can be recognized as any desired text. This is an end-to-end white-box attack, assuming   the attacker can obtain the structure and parameters of the identification system. The attack mode  is to send the speech directly to the ASR system, and it cannot attack in the air.

Given an original example $x$ and target text $t$, the optimization object   is
\begin{equation}
	Minimize \left | x\right |_{2}^{2}+c*l(x+\delta, t), such \; that \;dB_{x}(\delta)\leq \tau,
\end{equation}
where $c$  weighs whether to make the adversarial example closer to the original example or to make it easier to attack successfully, $l(\cdot, \cdot )$ is the loss function, and $dB_{x}(\delta)\leq \tau$  ensures that the perturbation  is not too large. To calculate the loss function requires a definite alignment $\pi$. The attack algorithm has two steps. An initial adversarial example is generated by CTC loss, which defines the current $\pi$. Fixing the current $\pi$,  an adversarial example with less perturbation is generated.

DeepSpeech is an end-to-end text-dependent ASR system based on a DNN \cite{Ref28} that has higher recognition performance than traditional methods, with an excellent effect  in  noisy environments. Experimental results   [14] have shown that   adversarial examples generated by the OPT method can enable DeepSpeech to output a specified text content with a 100\% attack success rate. The average disturbance size of the generated adversarial example is $-31dB$. The longer the length of specified text the more difficult it is to generate. The perturbation of the generation will also increase; on average, each extra character will increase the perturbation by $0.1dB$. If the text of the original example is longer, the adversarial example will be less difficult to generate.
\subsection{GA-based method}
\label{sec:2-2}
This method uses a genetic algorithm based on  a free gradient  to generate adversarial examples on command-dependent ASR systems.   Shown as Algorithm 1,  it uses  normal speech and a target command as   input, and   creates a group of candidate adversarial examples by adding random noise to the subset examples in a given speech segment. To minimize the impact of noise on human hearing perception, it is only added to the least significant bit (LSB) of the speech. The deterministic score of each population member is calculated according to the predicted score of the target label. Through the application of selection, crossover, and mutation,   next-generation adversarial examples are generated from the current generation. Members of the population with higher scores are more likely to be part of the next generation. Crossover is to mix pairs of population members to generate a new example and add it to the new population. Mutation adds random noise to the offspring with minimal probability before passing it on   to the next generation. The process is repeated  until a preset value is reached or the attack is successful.

\renewcommand{\algorithmicrequire}{\textbf{Inputs  :}} 
\renewcommand{\algorithmicensure}{\textbf{Output:}}
\begin{algorithm}[H]
	\caption{Genetic Algorithm Based on Adversarial Example Generation}
	\label{alg:1}
	\begin{algorithmic}[H]
		\Require
		Original example $x$;
		target label $t$
		\Ensure
		Targeted adversarial example $x_{adv}$
		\State $pop:=InitializePopulation(x)$
		\State $k_{iter}=0$
		\While{($k_{iter}<k_{max}$)}
		
		\State $scores:=ComputeFitness(pop)$
		\State $x_{adv}:=pop[argmax(scores)]$
		\If {$argmax \; f(x_{adv})=t$}
		\State break
		\EndIf
		\State $probs:=softmax(\frac{scores}{temp})$
		\State $pop_{next}:=$\{\}
		\For{$i:=1$ to $size$}
		\State $parent_{1},parent_{2}=randomChoice(pop, probs)$
		\State $child=Crossover(parent_{1}, parent_{2})$
		\State $pop_{next}:=pop_{next}\cup$\{$child$\}
		\EndFor
		\ForAll{$child$ of next $pop$} $Mutate(child)$
		\State $pop:=pop_{next}$
		\State $k_{iter}:=K_{iter}+1$
		\EndFor
		\EndWhile
		\State \Return $x_{adv}$
	\end{algorithmic}
\end{algorithm}

The command-dependent ASR system SpeechCommand \cite{Ref31}  recognizes a command label from speech, and is essentially a multi-classification network. Experimental results have shown that its success rate of attack   can reach 87\%.

%
%

\section{Devastation and detection on adversarial examples}
\label{sec:3}
We introduce devastation and detection on   examples generated by attack methods. Present methods for the defense of speech adversarial examples   modify the training process, change the structure of the network model, or add additional models. These operations can require much computation and training overhead. We propose the devastation and detection of speech adversarial examples based on the addition of random noise, discuss its influence, and  provide examples of devastation and detection methods.
\subsection{Devastation on adversarial examples} 
\label{sec:3-1}
Given the speech signal $x$,   attackers use the adversarial example generation algorithm to generate a local gradient in the input layer of the network structure, which is consistent with the size of the input signal $x$. We add  perturbation $\delta^{*}$ to   $x$ to generate  adversarial example

\begin{equation}
	x^{*}=x+\delta^{*}.
\end{equation}

Random noise $\hat{\delta}$ is added to the adversarial example, whose size   is consistent with   Gaussian noise, to generate adversarial example $\hat{x}^{*}$. Mixed noise $\hat{\delta}^{*}$ is the sum of the addition of disturbance $\delta^{*}$ and Gaussian noise $\hat{\delta}$,
\vspace{-10pt}
\begin{equation}
	\hat{x}^{*}=x^{*}+\hat{\delta},\hat{\delta}(x,\mu,\sigma)=\frac{1}{\sqrt{2\pi}\sigma}exp(-\frac{(x-\mu)^{2}}{2\sigma^{2}}),\hat{\delta}^{*}=\delta^{*}+\hat{\delta},
\end{equation}
where the average $\mu$   and standard deviation $\sigma$   can represent the intensity of noise added to the speech signal. Finally, the modified adversarial example can be equivalent to the addition of mixed noise $\hat{\delta}^{*}$ to the original speech,

\begin{equation}
	\hat{x}^{*}=x+\hat{\delta}^{*}.
\end{equation}

Because there is a certain direction when adding adversarial example perturbation, which is equivalent to the addition of purposeful disturbance to make the example close to the target class, these small perturbations play a great role in the discrimination of the model. Even if the input data change slightly, the final calculated value is closer to the distribution of the target class after a series of calculations. When the intensity   of of $\delta^{*}$ is similar to that of $\hat{\delta}$, or the strength   is greater, the superimposed noise $\hat{\delta}^{*}$ will lose the particularity of $\delta^{*}$ and    become ordinary noise, which will affect the purpose of adversarial perturbation, i.e., the adversarial example $\hat{x}^{*}$ will not be adversarial, and the devastation strategy will work.

After adding random noise $\hat{\delta}$ to the normal speech signal $x$,    we can obtain the modified normal speech, i.e., some useless values are added to the normal speech signal. If the intensity of noise is slight, the speech   sounds like the original. Even if the modified normal speech is put into the classifier, the result will not be greatly changed. When the intensity of $\hat{\delta}$ is small, the result is similar to   normal. In other words, the addition of small random noise has little effect on the recognition of normal examples, and the noise lacks a direction. As shown in Fig. 1, adding random noise to the normal example, the result will not be changed, but the adversarial example will lose the effect of the attack. Because the adversarial perturbation is devastated by the noise, the model discriminating the speech is affected by the input signal. Adversarial examples that carry a purposeful value can make a model misclassify the signal.
\begin{equation}
	\hat{x}=x+\hat{\delta}
\end{equation}

\begin{figure}[H]
	\centering
	\includegraphics[scale=0.5]{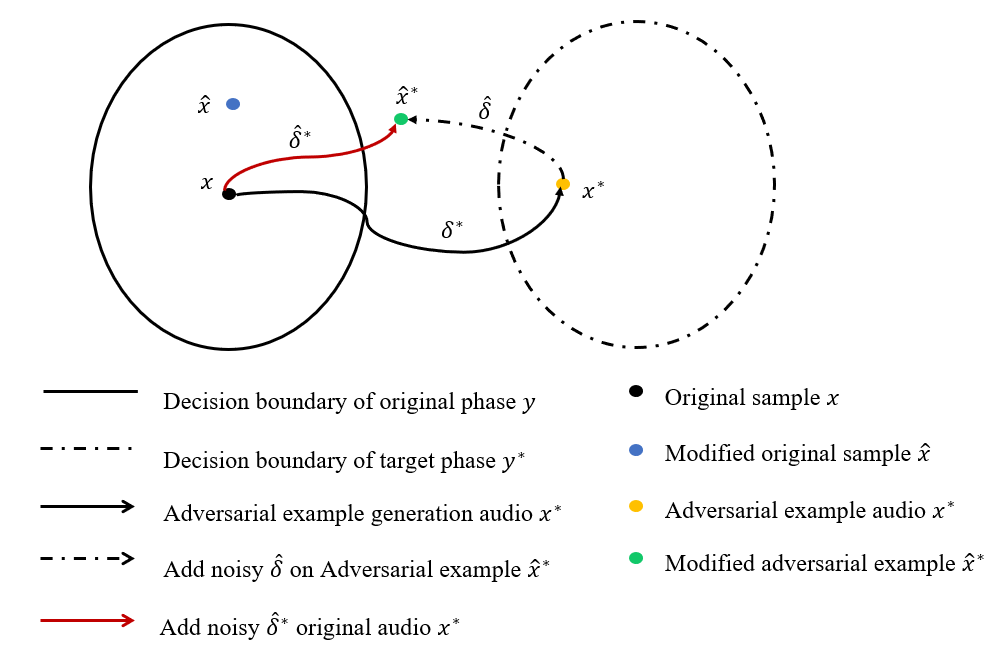}
	\caption{One single normal example $x$ adds perturbation, and different noise can change the recognition result of the ASR system. The normal example $x$ with perturbation $\delta^{*}$ will be changed to an adversarial example $x^{*}$, and adversarial example  $\delta^{*}$ with random noise $\hat{\delta}$ will be changed to  an unknown example $\hat{x}^{*}$.}
	\label{fig:pathdemo1}
\end{figure}

Using a normal example of TIMIT, and generating an adversarial example by OPT algorithm, random   and Gaussian noise are added to the normal   and   adversarial examples. Fig. 2 shows the spectrogram of the example. Table 1 shows the recognition results of examples in the DeepSpeech system. It can be concluded that  the addition of noise to the normal examples does not change the recognition results. However, the recognition result of the adversarial example   changes greatly, and is close to that of the normal example. This shows that the slight random noise will not affect the recognition result of the normal example, but it can invalidate the  adversarial example and become closer to the normal example. Therefore, we can use random noise to devastate the adversarial example without seriously affecting the normal example. According to these experimental results, we can further propose the detection method of adversarial examples.

\begin{table}[H]
	\centering
	\caption {DeepSpeech   examples:  normal and adversarial examples with added ordinary  and Gaussian random noise}
	\label{tab:chap:table_1}
	\resizebox{\textwidth}{15mm}{
		\begin{tabular}[c]{lll}
			\toprule[0.5pt]
			{Example} & {Recognition result}  \\
			\midrule[1pt]
			(a) normal  & she had her dark suiting greacy wash water all year  \\
			(b) adversarial  & this is an adversarial example  \\
			(c) normal  with random noise & she had yedark sutin greacy wash water all year  \\
			(d) adversarial  with random noise & he had regark suting greacy watch water all yer  \\
			(e) normal  with Gaussian noise & she had redark sutin greacy watch water all yer  \\
			(f) adversarial  with Gaussian noise & he had redark suvin greacy watch water all year  \\
			\bottomrule[0.5pt]
	\end{tabular}}
\end{table}

\begin{figure}[H]
	\centering
	
	\subfigure[normal]{
		\begin{minipage}[t]{0.5\linewidth}
			\centering
			\includegraphics[width=2in]{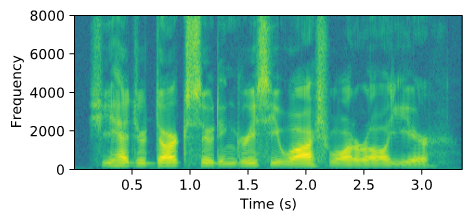}
		\end{minipage}%
	}%
	\subfigure[adversarial]{
		\begin{minipage}[t]{0.5\linewidth}
			\centering
			\includegraphics[width=2in]{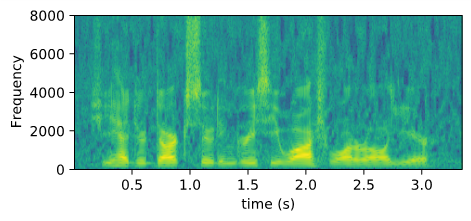}
		\end{minipage}%
	}%
	
	\subfigure[normal  with random noise]{
		\begin{minipage}[t]{0.5\linewidth}
			\centering
			\includegraphics[width=2in]{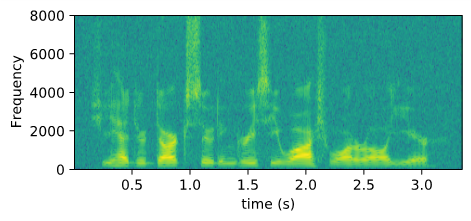}
		\end{minipage}
	}%
	\subfigure[adversarial with random noise]{
		\begin{minipage}[t]{0.5\linewidth}
			\centering
			\includegraphics[width=2in]{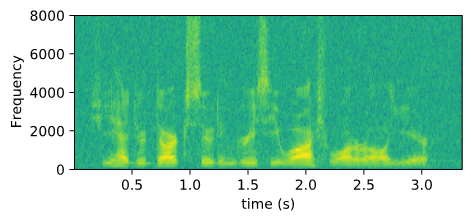}
		\end{minipage}
	}%

	\subfigure[normal with Gaussian noise]{
		\begin{minipage}[t]{0.5\linewidth}
			\centering
			\includegraphics[width=2in]{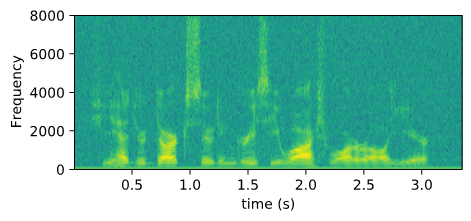}
		\end{minipage}
	}%
	\subfigure[adversarial with Gaussian noise]{
		\begin{minipage}[t]{0.5\linewidth}
			\centering
			\includegraphics[width=2in]{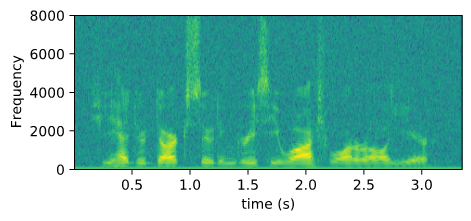}
		\end{minipage}
	}%
	
	\centering
	\caption{Spectrogram of normal  and adversarial examples with ordinary  and Gaussian noise}
\end{figure}

\subsection{Detection on adversarial examples} 
\label{sec:3-2}
The detection algorithm is motivated by the devastation algorithm. Given a speech sample $x$ (normal   or adversarial), we only need to add noise   before inputting it to the ASR system. For a normal example, because the low-intensity random noise will not affect the content of the speech, the recognition result of the ASR system will not change much. For an adversarial example, because the perturbation is added to the current speech, the  added random noise will devastate the particularity of the perturbation, and the recognition result of the ASR system will be totally different from the original. Therefore, according to the recognition results before and after adding noise, we can determine whether an example is adversarial.
\begin{figure}[H]
	
	\centering
	\includegraphics[scale=0.4]{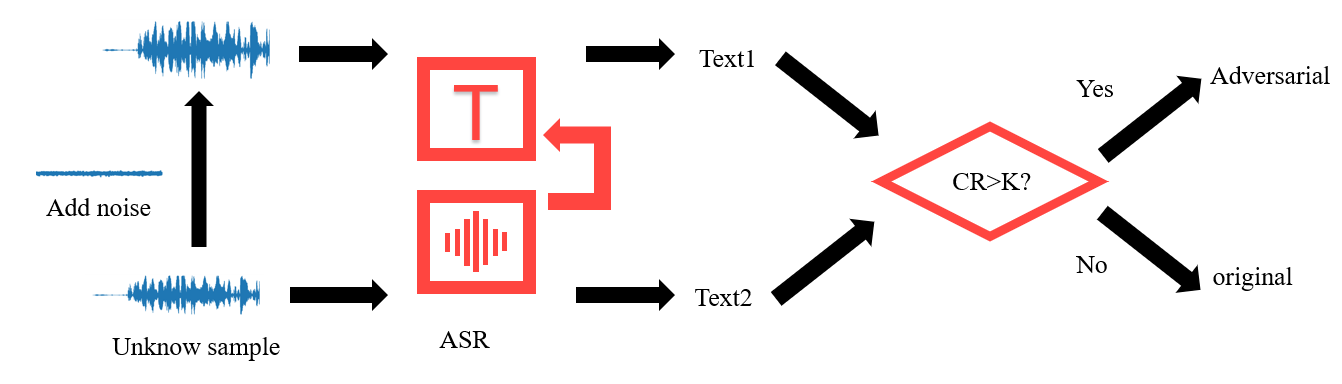}
	\caption{Flowchart of detection of adversarial examples by $CR$   before and after adding noise; the example is adversarial if $CR$ is greater than a threshold $K$.}
	\label{fig:pathdemo1}
\end{figure}

The   detection strategy determines whether   unknown speech is an adversarial example. Because the recognition result of such an example is more likely to be affected by random noise, we can use the change rate (CR) of the recognition result  after adding random noise to detect an unknown example. According to the   recognition results   in Table 1, the variation of the adversarial example is very large relative to a normal sample. As shown in Fig. 3, we can detect the adversarial example following   the process  shown as Algorithm 2.
\begin{algorithm}[H]
	\caption{Detection of   Adversarial Example}
	\label{alg:1}
	\begin{algorithmic}[H]
		\Require
		Unknown example $x$;
		\Ensure
		Result of   detection 
		\State /* Add   noise $\hat{\delta}$ on   sample */
		\State $\hat{x}$=x+$\hat{\delta}$
		\State $D=Dist(a,b)$
		\State /* Calculate change rate (CR) between $x$ and $\hat{x}$, $g(a)$ is recognition system */
		\State $L=g(x)$
		\State $CR=\frac{\min(D(g(\hat{x}),g(x)),L)}{L}$
		\If{CR$\textgreater$K}
		\State Example $x$ is   adversarial
		\Else
		\State  Example $x$ is  normal
		\EndIf
	\end{algorithmic}
\end{algorithm}

According to Algorithm 2, whether an unknown example is   adversarial   can be detected on the basis of not seriously devastating   the example. In terms of the previous change in the recognition rate of the example after adding noise,  the change beyond a certain threshold can indicate an adversarial example. From our experiments, it can be concluded that   random noise will not have much impact on a normal example.

\section{Experimental results}
\label{sec:4}
\subsection{Database}
\label{sec:4-1}
Our experimental data  included   text-    and command-dependent speech databases. The text-based speech database was made up of TIMIT and LibriSpeech, both with a sampling rate of 16 kHz and a bit depth of 16 bits. TIMIT is an acoustics-phoneme continuous speech corpus built by Texas Instruments, Massachusetts Institute of Technology, and SRI International.   The database includes 630 speakers from different parts of the United States,  70\%   male, and mostly adult and white. Each   participant spoke 10  sentences, and a total of 6,300 examples were obtained, all   manually tagged at the phoneme level.   LibriSpeech \cite{Ref29} is a corpus of about 1000 hours of English pronunciation from audiobooks from the LibriVox project.   In our experiments, we downloaded a test-clean dataset and used the first 100 examples.

The SpeechCommand Dataset (SpeechCommands) from Google contains 105829 speech files, each  consisting of 35   words. The sampling frequency   is 16 kHz, the bit depth is 16 bits, the duration is near 1 second, and the format is WAV. The dataset includes 2618   participants who were asked to say 35  words,   each   participating only once. They had 1.5 seconds to read out each  word, at one-second intervals. Examples with no sound content or whose speech content was different from the words were deleted. Each segment was checked manually to delete   speech content that was different from the given words. Recordings are in OGG format, and files   smaller than 5 KB were deleted. Speech signals were converted to WAV format   normalized to [-1.0, 1.0], and     files whose average values of the normalized speech signal were less than 0.004 were deleted.
\subsection{Experimental setup}
\label{sec:4-2}
We used the GA and OPT attack methods (section 3.2) to generate   adversarial examples. For the text-dependent ASR system, we chose classical DeepSpeech\cite{Ref28} as our target system, and OPT   to generate adversarial examples to attack the system. TIMIT, LibriSpeech, and CommandVoice were the speech databases. For the command-dependent ASR system, SpeechCommand was the target, and the GA-based method was chosen to generate adversarial examples. For   added noise,  $\mu$ had the range (10, 30, 50, 70, 100, 200, 500), and   $\sigma$ was in  the range (10, 30, 50, 70, 100, 200, 500).

Defense methods on adversarial examples were evaluated  differently for text- and command-dependent ASR systems. The defense method of text-dependent adversarial examples is illustrated from two aspects: (1) the effect on   adversarial examples was to be as large as possible, and we used the similarity of   recognition results    after adding noise $SR_{adv}$; and (2) the impact on normal examples was to be as small as possible, and we again measured it by the similarity of recognition results of normal examples   after adding noise $SR_{benign}$. Similarity is the matching ratio between the initial recognition result of a sample     after adding noise. We calculated
\begin{equation}
	SR_{benign}=\frac{D(T(x_{benign}),y)}{D(g(x_{benign}),y)}, SR_{adv}=\frac{D(T(x_{adv}),y)}{D(g(x_{adv}),y)},
\end{equation}
where $x_{benign}$ is a normal example, $x_{adv}$ is an adversarial example, $y$ is the real text, $D(\cdot ,\cdot )$ is the distance function, and the editing distance proposed by Levenshtein et al. \cite{Ref30}   represented the input transformation function (e.g., downsampling, quantization, local smoothing, compression).

The defense effect of command-dependent ASR system adversarial examples is illustrated from two aspects: (1) the impact on the adversarial example should be large, as measured by the change of the average attack success rate $ASV_{avg}$   after adding noise; and (2) the impact on a normal example should be small, as measured by the change of the recognition accuracy $ACC$   after adding noise.

In the experiment, we calculated 
\begin{equation}
	ASR_{avg}=\frac{\sum_{x^{*}}^{X^{*}}(g(M(x^{*}))==y^{*})}{n^{*}},
\end{equation}
where $X^{*}$ is the adversarial example set generated from original example set $X$, $n^{*}$ is the number of examples in $x^{*}$, $x^{*}$ is a single example in $X^{*}$, and $y^{*}$ is a label for attacking $x^{*}$. A  lower $ASR_{avg}$ on adversarial examples indicates better performance of the method.

The recognition accuracy is the ratio of the number of correctly recognized speech examples   to the total,
\begin{equation}
	ACC=\frac{\sum_{x}^{X}(g(M(x))==y_{0})}{n},
\end{equation}
where dataset $X$ has $n$  examples, and $x$ is an example whose real label is $y_{0}$. The closer   ACC is  before and after adding noise, the smaller the influence of the defense method, and the better the defense performance.

\subsection{Results and discussion}
\label{sec:4-3}
We show the effect of   devastation on   text-dependent adversarial examples, and   detection results on   command-dependent adversarial examples. 
\subsubsection{Results of devastation on adversarial examples}
\label{sec:4-3-1}
We    explore the effect of the intensity of ordinary and Gaussian noise on the experimental results, and compare our method with other advanced methods.

Table 2   compares the similarity of normal examples (NEs) and adversarial examples (AEs) with different intensities of ordinary noise   after processing. With   increasing noise intensity, the similarity of NEs decreases gradually, and that of AEs decreases greatly. When the intensity of the noise  is close to 50, the similarity of AEs is 0\% on the TIMIT and LibriSpeech databases, and   12\% on the CommonVoice database. When the noise intensity is higher than 50, the similarity of AEs is unchanged. Therefore, the appropriate general noise intensity is about 50.
\begin{table}[H]
	\centering
	\caption {Similarity results of ordinary noise addition on three databases with seven   noise intensities; the impact is large on adversarial examples and small on normal examples.}
	\label{tab:chap:table_1}
	\begin{tabular}{ccccccc}
		\hline
		\multirow{3}{*}{Parameter} & \multicolumn{6}{c}{Datasets}                                                                                                                              \\ \cline{2-7} 
		& \multicolumn{2}{c}{TIMIT}                         & \multicolumn{2}{c}{LibriSpeech}                   & \multicolumn{2}{c}{CommonVoice}                   \\
		& \multicolumn{1}{l}{NEs} & \multicolumn{1}{l}{AEs} & \multicolumn{1}{l}{NEs} & \multicolumn{1}{l}{AEs} & \multicolumn{1}{l}{NEs} & \multicolumn{1}{l}{AEs} \\ 
		\midrule[1pt]
		10                         & \textbf{96.81}          & 81.04                   & \textbf{99.68}          & 96.06                   & \textbf{97.41}          & 79.27                   \\
		50                         & 88.50                   & \textbf{0}              & 98.38                   & \textbf{0}              & 92.70                   & 12.00                   \\
		100                        & 80.18                   & 0                       & 97.20                   & 0                       & 88.20                   & \textbf{0}              \\
		200                        & 62.85                   & 0                       & 94.48                   & 0                       & 83.56                   & 0                       \\
		300                        & 48.29                   & 0                       & 92.36                   & 0                       & 78.56                   & 0                       \\
		400                        & 37.58                   & 0                       & 90.40                   & 0                       & 75.49                   & 0                       \\
		500                        & 29.63                   & 0.43                    & 87.66                   & 0                       & 71.29                   & 0                       \\ \hline
	\end{tabular}
\end{table}

Table 3   compares the similarity of NEs and AEs with different intensities of Gaussian noise. With   increasing   noise intensity, the similarity of NEs decreases gradually, and that of AEs decreases greatly. When the noise intensity is close to 50, the similarity of   AEs of the three databases is almost zero. When the noise intensity is higher than 50, the similarity of AEs is unchanged. On the TIMIT database, when the noise intensity is higher than 300, the similarity of AEs increases slightly, perhaps because the perturbation   is small,   the noise intensity of 300 is already higher than that of the perturbation, and the influence of adding noise intensity becomes small. Therefore, the appropriate intensity of Gaussian noise is about 50. Comparison with the results of Table 2 shows that  ordinary noise has a slightly better result than  Gaussian noise.

Table 4   compares the similarity results of NEs and AEs with different methods. For the method of Kwon,  the similarity of the 8-bit reduction on AEs on the three databases is   low, as is the similarity of  NEs in the TIMIT database. Hence, this method is not suitable for all databases. A low-pass filter performs well on both normal   and adversarial examples. The similarity on   AEs of silence removal is high. For the Yang method, the similarity of the four processing samples on the three databases is high, and the similarity of NEs of Quan-512 is the lowest. For our method, the results of random noise-50 are better than  those of other methods, and   NEs have  high similarity, but the similarity of   AEs on the CommonVoice database is nonzero. With Gaussian noise-50, the similarity of AEs in three databases is zero. Compared with low-pass filtering, the result is lower on  TIMIT, higher on  LibriSpeech, and similar on CommonVoice. Our method has an overall better effect than the others.

\begin{table}[H]
	\centering
	\caption {Similarity result of noise addition on databases with seven noise intensities; the impact is large on adversarial examples and small on normal examples.}
	\label{tab:chap:table_1}
	\begin{tabular}{ccccccc}
		\hline
		\multirow{3}{*}{Parameter} & \multicolumn{6}{c}{Datasets}                                                                                                                              \\ \cline{2-7} 
		& \multicolumn{2}{c}{TIMIT}                         & \multicolumn{2}{c}{LibriSpeech}                   & \multicolumn{2}{c}{CommonVoice}                   \\
		& \multicolumn{1}{l}{NEs} & \multicolumn{1}{l}{AEs} & \multicolumn{1}{l}{NEs} & \multicolumn{1}{l}{AEs} & \multicolumn{1}{l}{NEs} & \multicolumn{1}{l}{AEs} \\ \midrule[1pt]
		10                         & \textbf{96.81}          & 81.04                   & \textbf{99.68}          & 96.06                   & \textbf{97.41}          & 79.27                   \\
		50                         & 88.50                   & \textbf{0}              & 98.38                   & \textbf{0}              & 92.70                   & 12.00                   \\
		100                        & 80.18                   & 0                       & 97.20                   & 0                       & 88.20                   & \textbf{0}              \\
		200                        & 62.85                   & 0                       & 94.48                   & 0                       & 83.56                   & 0                       \\
		300                        & 48.29                   & 0                       & 92.36                   & 0                       & 78.56                   & 0                       \\
		400                        & 37.58                   & 0                       & 90.40                   & 0                       & 75.49                   & 0                       \\
		500                        & 29.63                   & 0.43                    & 87.66                   & 0                       & 71.29                   & 0                       \\ \hline
	\end{tabular}
\end{table}

\begin{table}[H]
	\centering
	\caption {Similarity results of state-of-the-art   and proposed methods; ours performs better than the methods of Kwon and Yang}
	\label{tab:chap:table_1}
	\begin{tabular}{cccccccc}
		\toprule[0.25pt]
		\multirow{3}{*}{Method}   & \multirow{3}{*}{Type} & \multicolumn{6}{c}{Datasets}                                                                   \\ \cline{3-8} 
		&                       & \multicolumn{2}{c}{TIMIT}  & \multicolumn{2}{c}{LibriSpeech} & \multicolumn{2}{c}{CommonVoice} \\
		&                       & NEs           & AEs        & NEs              & AEs          & NEs              & AEs          \\ \midrule[1pt]
		\multirow{2}{*}{Random}   & Random noise-50       & \textbf{88.5} & \textbf{0} & \textbf{98.38}   & \textbf{0}   & \textbf{92.70}   & 12.00        \\
		& Gaussian noise-50     & 82.81         & 0          & 97.35            & 0            & 90.07            & \textbf{0}   \\
		\multirow{3}{*}{Kwon \cite{Ref27}} & 8-bit reduction       & 59.26         & 0          & 93.91            & 0            & 82.08            & 0            \\
		& Low-pass filtering    & 87.28         & 0          & 93.95            & 0            & 90.74            & 0            \\
		& Silence removal       & 73.64         & 22.43      & 94.35            & 0            & 83.56            & 09.09        \\
		\multirow{4}{*}{Yang \cite{Ref26}} & Downsampling         & 85.54         & 20.48      & 93.37            & 19.53        & 87.91            & 14.86        \\
		& Smoothing              & 77.61         & 20.44      & 85.99            & 19.03        & 82.19            & 14.97        \\
		& Quan-256              & 77.61         & 21.93      & 96.73            & 21.08        & 88.44            & 20.00        \\
		& Quan-512              & 59.30         & 14.28      & 93.86            & 19.39        & 81.42            & 18.38        \\ \hline
	\end{tabular}
\end{table}

\subsubsection{Results of detection on adversarial examples}
\label{sec:4-3-2}
We discuss the results of   AE detection on the command-dependent ASR system. We    explore the effect of the intensity of ordinary and Gaussian noise, and compare our method with others.

Table 5 shows the changes of $ASR_{avg}$ and $ACC$ of AEs with different intensities of ordinary noise. With   increasing   noise intensity, both measures show a downward trend, with $ASR_{avg}$ decreasing more than $ACC$. When the noise intensity is higher than 100,   $ASR_{avg}$ is  less than 10\%. In practical scenes, a tradeoff should be made between $ASR_{avg}$ and $ACC$. When the change of $ACC$ is small,   noise with larger intensity should be selected to make   $ASR_{avg}$ as small as possible.

\begin{table}[H]
	\centering
	\caption{Results of different intensities of ordinary random noise of $ASR_{avg}$ and $ACC$}
	\begin{tabular}{ccc}
		\toprule[0.5pt]
		{Parameters} & {$ASR_{avg}$ (\%)} & {$ACC$ (\%)} \\
		\midrule[1pt]
		10  & 52.54 & \textbf{93.80} \\
		30  & 32.89 &\textbf{93.80}   \\
		50  & 21.96 & 93.40            \\
		70  & 16.29 & 93.60             \\
		100 & 10.62 & 94.00              \\
		200 & 3.98  & 92.00               \\
		500 & \textbf{1.80} & 88.80        \\
		\bottomrule[0.5pt]
	\end{tabular}
\end{table}

\begin{table}[H]
	\centering
	\caption{Results of different intensities of Gaussian random noise of $ASR_{avg}$ and $ACC$}
	\begin{tabular}{ccccccccc}
		\toprule[0.5pt]
		{Parameters} & {$ASR_{avg}$ (\%)} & {$ACC$ (\%)} \\
		\midrule[1.1pt]
		10  & 42.60 & \textbf{94.00} \\
		30  & 20.62 & \textbf{94.00}  \\
		50  & 11.64 & 93.20            \\
		70  & 7.92  & 93.60             \\
		100 & 4.80  & 92.60              \\
		200 & 2.13  & 90.80               \\
		500 & \textbf{1.75}  & 86.00       \\
		\bottomrule[0.5pt]
	\end{tabular}
\end{table}

Table 6 shows the changes of   $ASR_{avg}$ of AEs and $ACC$ of NEs under   different intensities of Gaussian noise. Similar to the results of ordinary noise, both $ASR_{avg}$ and $ACC$ show a downward trend with the increase of noise intensity. Compared to Table 5, with the same noise intensity,   $ASR_{avg}$  and   $ACC$ under Gaussian noise are both smaller than under ordinary noise. We can conclude that for a better defense effect, we can add Gaussian noise, and if we reduce the recognition accuracy of NEs, we can choose ordinary noise.
\begin{table}[H]
	\centering
	\caption {Comparison of $ASR_{avg}$ and $ACC$   with   state-of-the-art   and   proposed methods; our method outperforms  the methods of Kwon and Yang.}
	\label{tab:chap:table_1}
	\begin{tabular}{cccc}
		\toprule[0.5pt]
		Method  & Type & $ASR_{avg}$ (\%) & $ACC$ (\%) \\ 
		\midrule[1pt]
		Without defense   &  -    &   83.81  &  95.00   \\ 
		\multirow{2}{*}{Random} &  Random noise-200    &  03.98   &  \textbf{92.00}  \\
		&  Gaussian noise-200    &  \textbf{02.13}   &   90.80  \\ 
		\multirow{3}{*}{Kwon \cite{Ref27}} &  8-bit reduction    &   02.82  &   92.00  \\
		&   8-bit reduction   &   28.78  &  90.60   \\
		&   Silence removal   &   09.22  &   85.00  \\ 
		\multirow{4}{*}{Yang \cite{Ref26}} &  Downsampling    & 10.22    &  91.60   \\
		&   Smoothing   &   20.78  &   92.00  \\
		&    Quan-256  &   29.97  &  90.20   \\
		&   Quan-512   &  07.84   &   89.00  \\ 
		\bottomrule[0.5pt]
	\end{tabular}
\end{table}
Table 7 shows the $ASR_{avg}$ of AEs with different defense methods and $ACC$ of NEs. $ASR_{avg}$ is calculated from the matrix of attack success rates of   samples of the corresponding defense methods in Fig. 4. We compare the results of the appropriate noise intensity from the experiments in Tables 5 and   6 to those of other methods. It can be seen that our method has a higher $ACC$ than others, while ensuring a lower $ASR_{avg}$. The $ASR_{avg}$ of Gaussian noise with intensity 200 is the smallest, and the $ACC$ of ordinary noise with intensity 200 is the largest. Kwon's 8-bit reduction also has a good effect, but its result on the text-dependent ASR system is poor. The $ASR_{avg}$ of other methods is high.

\begin{figure}[H]
	
	\centering
	\includegraphics[scale=0.4]{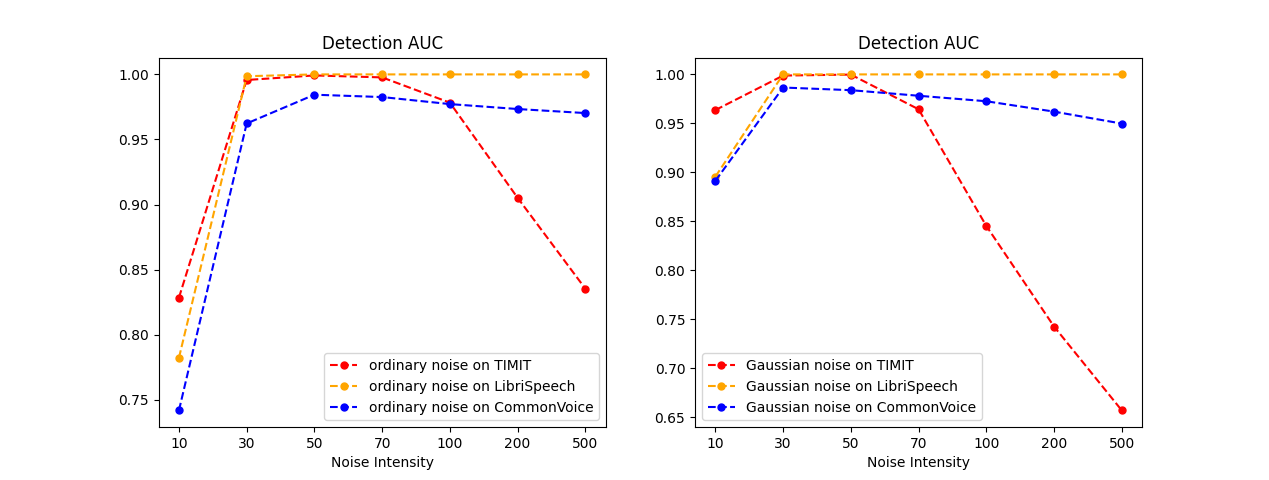}
	\caption{Detection AUC of AEs. Under  appropriate noise, the AE can be detected with high probability. As the intensity of noise increases, the effect on   LibriSpeech is good.}
	\label{fig:pathdemo1}
\end{figure}

The experimental results show that   noise addition has a great impact on   AEs, and reduces the success rate of attacks. For   NEs, the impact is   small, and the recognition accuracy is reduced very little. Fig. 5 shows the AUC of the AE detection rate under  different kinds and intensities of noise. Under the devastation of appropriate noise, the $CR$ fed back by the ASR system can well show whether a sample is adversarial. Hence, the category of examples can be calculated by the $CR$ after adding noise.   $CR$ is small for the NE and large for the AE. Without affecting the judgment of   normal examples by the ASR system,   AEs can be distinguished with great accuracy.

In conclusion,   noise addition   is confirmed to devastate and detect   AEs,  with  little cost for NEs. The experimental results are better than those of the advanced method.

\section{Conclusions}
\label{sec:5}
We proposed an algorithm for speech adversarial example devastation and detection based on the addition of random noise. After adding an appropriate intensity of noise to an adversarial example, its perturbation     becomes the sum of the original perturbation and random noise. Noise devastates the original perturbation, and it loses its particularity, so the adversarial example after adding noise will lose the attack effect. We chose   ordinary random noise and Gaussian noise. In experiments, we used the text-dependent  DeepSpeech ASR system with the OPT attack method, and the   command-dependent CommonVoice ASR system  with the GA-based attack method. When choosing an appropriate noise intensity and   type, our method was better than those of Kwon and Yang, and  ordinary random noise had a slightly better effect than Gaussian noise. Although our proposed method of random noise addition can devastate adversarial examples, it also affects the recognition results of normal examples, and the appropriate intensity of noise is related to the perturbation intensity of adversarial examples. In our future work, we will formulate   methods to overcome these shortcomings so as to achieve better results.

\begin{figure}[H]
	\centering
	
	\subfigure[without defense]{
		\begin{minipage}[t]{0.4\linewidth}
			\centering
			\includegraphics[width=1.4in]{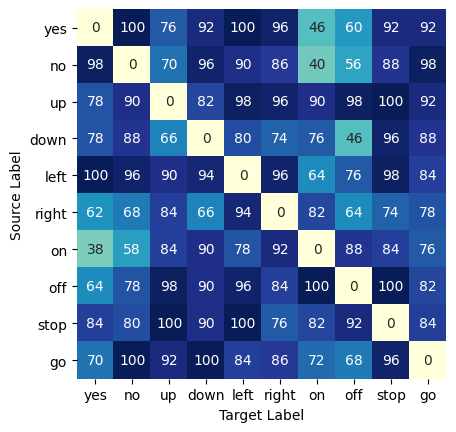}
		\end{minipage}%
	}%
	\subfigure[random noise-200]{
		\begin{minipage}[t]{0.5\linewidth}
			\centering
			\includegraphics[width=1.4in]{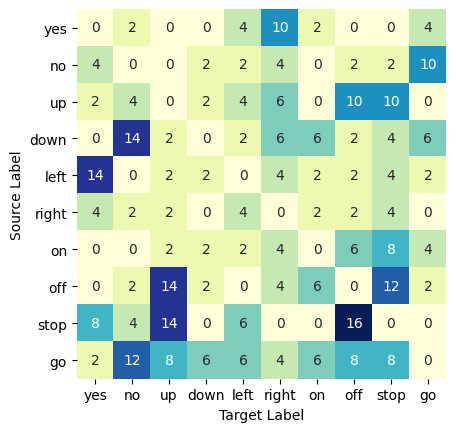}
		\end{minipage}%
	}%
	
	\subfigure[Gaussian noise-200]{
		\begin{minipage}[t]{0.4\linewidth}
			\centering
			\includegraphics[width=1.4in]{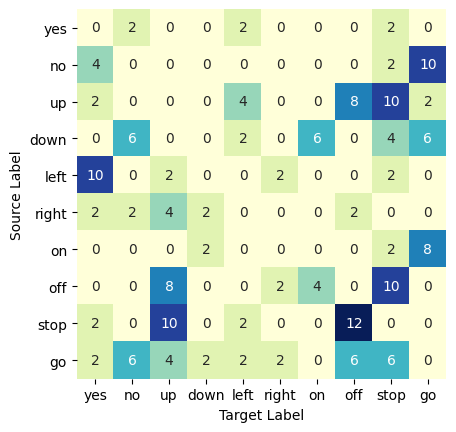}
		\end{minipage}
	}%
	\subfigure[8-bit reduction]{
		\begin{minipage}[t]{0.5\linewidth}
			\centering
			\includegraphics[width=1.4in]{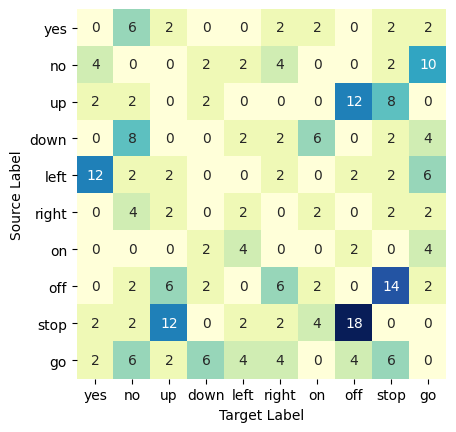}
		\end{minipage}
	}%

	\subfigure[low-pass filtering ]{
		\begin{minipage}[t]{0.4\linewidth}
			\centering
			\includegraphics[width=1.4in]{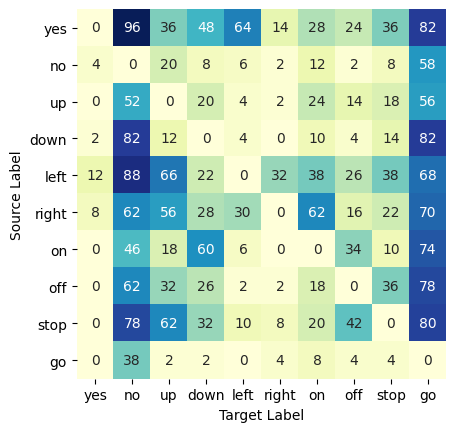}
		\end{minipage}
	}%
	\subfigure[silence removal]{
		\begin{minipage}[t]{0.5\linewidth}
			\centering
			\includegraphics[width=1.4in]{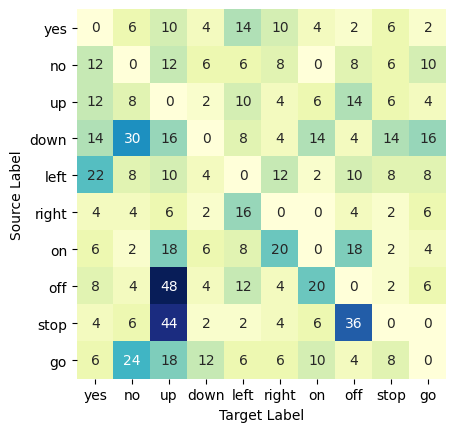}
		\end{minipage}
	}%
	
	\subfigure[downsampling]{
		\begin{minipage}[t]{0.4\linewidth}
			\centering
			\includegraphics[width=1.4in]{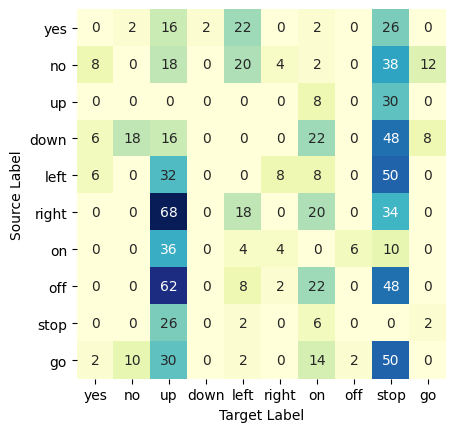}
		\end{minipage}
	}%
	\subfigure[smoothing]{
		\begin{minipage}[t]{0.5\linewidth}
			\centering
			\includegraphics[width=1.4in]{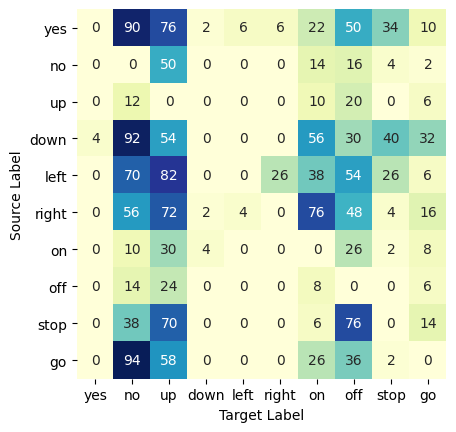}
		\end{minipage}
	}%

	\subfigure[Quan-256]{
		\begin{minipage}[t]{0.4\linewidth}
			\centering
			\includegraphics[width=1.4in]{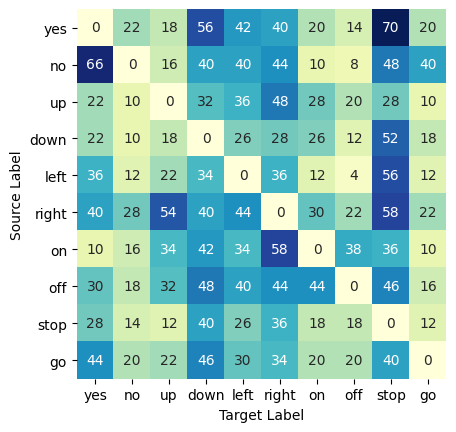}
		\end{minipage}
	}%
	\subfigure[Quan-512]{
		\begin{minipage}[t]{0.5\linewidth}
			\centering
			\includegraphics[width=1.4in]{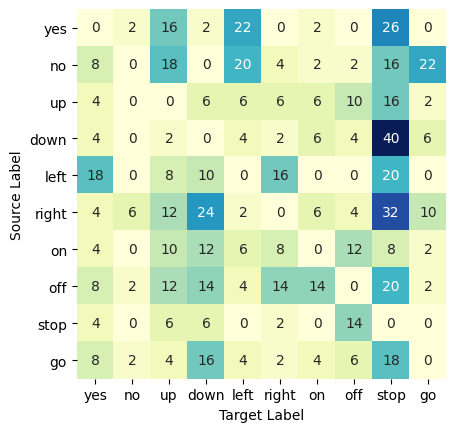}
		\end{minipage}
	}%
	
	\centering
	\caption{Evaluation metrics comparing state-of-the-art methods with ours.}
\end{figure}



\section*{Acknowledgements}
This work was supported by the National Natural Science Foundation of China (Grant No. 6217011361, U1736215, 61901237), Zhejiang Natural Science Foundation (Grant No. LY20F020010), Ningbo Natural Science Foundation (Grant No. 202003N4089) and K.C. Wong Magna Fund in Ningbo University.

\end{document}